# On the Consilience between QBism and Phenomenology

Hans Christian von Baeyer, College of William and Mary

hcvonb@wm.edu

*Tout notre raisonnement se réduit à céder au sentiment.*
*(All our reasoning boils down to yielding to feeling.)*
  -- Blaise Pascal

**Abstract**

Two decades after its creation, the interpretation of quantum mechanics called QBism is entering a new phase. Since it shares a personalist, subjective world-view with phenomenology, the philosophical study of human experience, there is a growing interest in the relationship between the two subjects. I call attention to the little-known philosopher Samuel Todes, whose phenomenology focused on the essential role of the human body in our understanding of the world. After reviewing this radical proposal, and illustrating it with some examples, I recommend it as an interesting mediator between the communities of physicists and phenomenologists. In addition, I argue that it may prove useful for promoting the public understanding of QBism.

## 1. Introduction

The study of the *interpretation of quantum mechanics* (IQM) is poised to enter a new epoch. The story of IQM began with the creation of quantum mechanics itself in 1925/1926, but its appeal as a research topic was usurped by the theory's unprecedented successes in explaining experiments and suggesting new devices. Around the year 2000, amid the rapid development of *quantum information theory*, novel information-based IQMs appeared. Among them was QBism, which here serves as an exemplar of a whole class of different approaches. The fundamental novelties of QBism[1] were its substitution of personalist Bayesianism for frequentist probability, and its recognition that quantum measurement outcomes were not available "for all the world to see," but rather of the nature of personal experiences. Now, twenty years on, the third epoch in the IQM story focuses on the fundamental role of perception, as opposed to concept formation, in theoretical physics (both quantum and classical). Since phenomenology is, crudely speaking, the study of perception, students of IQM should broaden their horizons to include this philosophical and psychological topic -- as indeed many of the founders of quantum mechanics thought a century ago. For popularizers of physics, including myself, this poses a triple challenge. First, we must learn a modicum of the language of philosophy. Then we must urge our physicist colleagues to temper their disdain for philosophy. Finally, we must marshal the most persuasive philosophical arguments in support of the relevance of phenomenology for QBism.

## 2. Seeking consilience between QBism and Phenomenology

At the beginning of the 21st century, the late biologist Edward O. Wilson re-introduced the felicitous 19th century word *consilience* into the vocabulary of the two-cultures debate. He set the stage in lofty terms: "The greatest enterprise of the mind has always been and always will be the attempted linkage of the sciences and humanities."[2] The Oxford English Dictionary, for its part, defines consilience as "Agreement between the approaches to a topic of different academic subjects, especially [of] science and the humanities." QBism was developed by physicists, and has attracted the attention of philosophers, particularly phenomenologists. This volume, together with its predecessor[3], offers a timely contribution to the effort of finding consilience between the two approaches.

Perhaps the most profound attempt to explore the connection between philosophy and physics, "between soul and matter", was proposed by Wolfgang Pauli under the rubric *Psychophysics*.[4] Sadly, he got distracted from the project, and died, much too early, before he got back to it. Another pioneer of quantum mechanics, Erwin Schrödinger, placed the origin of the clash between science and the humanities close to the beginning of Western philosophy, two and a half millennia ago, by quoting Democritus. After describing the Greek atomic hypothesis, Schrödinger continued:

> "Yet Democritus at the same time realized that the naked intellectual construction which in his world-picture had supplanted the actual world of light and color, sound and fragrance, sweetness, bitterness and beauty, was actually based on nothing but the sense perceptions themselves which had ostensibly vanished from it. In fragment D125, taken from Galen and discovered only about fifty years ago, he introduces the intellect (dianoia) in a contest with the senses (aesthesis). The former says: 'Ostensibly there is color, ostensibly sweetness, ostensibly bitterness, actually only atoms and the void'; to which the senses retort: 'Poor intellect, do you hope to defeat us while from us you borrow your evidence? Your victory is your defeat.' *You simply cannot put it more briefly and clearly.*"[5] [My emphasis.]

The whimsical Fragment D125 is a reminder of the obvious and fundamental realization that everything that enters our individual minds does so via the senses. Just as Descartes's c*ogito* symbolizes the logical relationship between thought and existence, Democritus's little gem encapsules the tension between physics and phenomenology. The relaxation of the tension implies consilience.

In modern times, Democritus's complaint against the intellect has been sharpened by science. Many sense experiences have indeed been linked to well-understood physical concepts: color to light frequency, warmth to molecular motion, sound to material waves, pressure to interatomic forces, etc. But a bewildering variety of optical illusions, ranging from the rainbow to subtle computerized effects, reminds us that many materialistic "explanations'' of sense experiences are much more problematic than what we learned in school.[6] Furthermore, conditions such as *synesthesia*, the involuntary confusion of different senses, which has no basis in physics, hint at the entangled map of cognitive pathways that awaits its future cartographers.

An accessible, concrete example of the power of consilience between philosophy and science is the story of the *Molyneux Question,*[7] which was debated by the philosophers Locke, Berkeley, Hume, and others, before turning up recently in a lab at MIT and an eye clinic in New Delhi.[8] The question concerns the perception of the shape of a solid object, a feat that can be achieved by two different senses. The concept of shape *per se* does not seem to be very problematic. Nevertheless, in 1688 William Molyneux wrote to Locke, asking, in effect: "Would a blind subject, on suddenly gaining sight, be able to recognize an object previously known only by touch?" For centuries even the phrasing of this question provoked controversy. Today, sketchy experimental evidence supports Locke's answer, which was NO. However, after a recovery period that could be as short as a few days, the newly-sighted were able to reconcile what they saw with what they had felt earlier. For 333 years Molyneux's trenchant question has continued to fascinate philosophers and scientists alike.

In the case of the Molyneux Question, the arc of the story stretched from philosophy to science. With respect to QBism the arrow points in the opposite direction: Initiated by physicists, QBism is now attracting the attention of philosophers. The two communities don't always speak the same language. In the mind/matter discussion, for example, here are some of the dichotomies in use:

> **Greek atomists:**       Intellect / Senses
> **The lay public:**       Thinking / Feeling
> **Physicists**:                Theory / Experiment
> **Phenomenologists:**  Conception / Perception

Physicists tend to focus on the third line, philosophers on the fourth. Rather than favoring one definition over another, in the spirit of consilience it will be necessary to develop a language comfortable to both communities. Theory and Conception, for example, refer to the **Imagination,** Experiment and Perception to human **Experience**. The meanings of the terms, however, overlap substantially. Painstaking analysis of abstract concepts, a core business of philosophy, has already borne fruit for QBists: It was the examination of the meaning of *probability* that inspired Quantum Bayesianism, the precursor of QBism. Conversely, QBism has contributed to the unraveling of the philosophical problem of the meaning of time.[9] Here I am not recommending any specific nomenclature, but suggesting that clarifying the terms of the conversation among QBists and Phenomenologists is an essential first step in promoting understanding.

### 3. Samuel Todes: A forgotten philosopher of the Human Body

In 1963 Samuel Todes (1927-1994)[10] submitted his doctoral dissertation in philosophy, titled *The Human Body as Material Subject of the World*, to Harvard University. Although he successfully continued his academic career, his thesis[11] was not published until almost thirty years later as part of a series of distinguished Harvard dissertations, including those of luminaries such as Goodman, Davidson, Putnam, and Quine. It would be another decade before it appeared in a more accessible form under the less daunting title *Body and World.*[12] In keeping with the simplification of this title, I will avoid the controversial technical terms *subject body* and *body subject*.[13]

In an essay dated 1993, a year before his death, Todes summarized his radical program essentially as follows: **All sense we can make, whether of what is actually perceivable, or possible as conceivable, draws on our sense of our [active] body, and is unintelligible without reference to it.**[14]

In his book, Todes points to what he calls a common, fundamental error: *the misinterpretation of the human body as merely a material thing in the world*.[15] Instead, he suggests that: "The human body is not only a material thing found in the midst of other material things in the world, but it is also, and moreover thereby, that material thing whose capacity to move itself generates and defines the whole world of human experience in which any material thing, including itself, can be found."

To illustrate his take on the relationship between perception and conception, he used the metaphor of a building: "There are two levels of objective experience: the ground floor of perceptually objective experience; and the upper storey of imaginatively objective experience (of which theoretical understanding is one kind), which presupposes for its objectivity(i.e. for its dependability as living quarters ) that the ground floor unto which it is built is itself on firm foundations."[16] The firm foundation is our individual sense of our own living body.

As physicist, I would suggest the related metaphor of a Cartesian frame of reference; a foundation is, after all, under the frame of a house. Furthermore, Todes's own word *reference* in the bolded quote above brings to mind the idea of a reference frame. In my image, an understanding of the world in terms of abstract concepts resembles the graph of a mathematical function which only makes sense with respect to a frame of reference whose origin, axes, and scale must be specified before the function itself. Todes proposes that the reference frame is defined by the perceptions of the human body in a manner I will try to illustrate.

Todes's choice of the body as the source of an immutable certainty on which to build philosophy is as surprising as it is bold. Nevertheless, it makes sense to me! Descartes's cogito, for example, worked for him, but although I can learn what he thought, I have no way of knowing how he thought. But I think I am safe in assuming that his body's perceptive mechanisms were similar to mine. Todes explains:

> "Modern evolutionary biology supports the contention that mankind has not evolved over historical time—that is, in the last five or ten thousand years—and makes it plausible that the species specific character of the human body may condition all human experience within some broad but definite limits. A phenomenological analysis, I will argue, justifies our current biological belief by demonstrating how our sense of our body plays a basic formative role in our making sense of everything. And not only in a culturally bound way. For historical knowledge, travel, reading, and imagination show that we can make *some* sense of different cultures. But the level at which our sense of our body enters into our sense-making is such that experience shorn of that body-sense is *unintelligible*. We cannot imagine having any sense at all of things of that type.
>
> I will try to show this for perceptual objects, which *actually* present themselves to us, and for ideas or concepts, which present to ourselves as forms of *possibility*." [Italics in the original.][17]

Before I provide concrete examples of what Todes is suggesting, I want to mention why his proposal appeals to me. As the problems of the world multiply and intensify, science is not only an important tool for defining them, but also one of the most reliable sources of solutions. Before such solutions can be implemented, however, scientists must first persuade politicians and the general public to accept the findings of science. To this end, a growing community of writers, reporters, and teachers is using all available means of communication in a grand enterprise of *public understanding of science.* Unfortunately, their task, which is mine too, grows increasingly difficult as science reaches unimaginable realms of vastness, minuteness and complexity. Sometimes I despair of ever reaching the elusive "informed public."

The first obstacle is scale. Quantum mechanics, for example, which is poised to reach into our lives in the immediate future via quantum computing and quantum cryptography, possibly even quantum currency, owes its unfortunate reputation for weirdness in large part to the unimaginable chasm between the typical dimensions of our world and those of the quantum world. If QBism, with the help of phenomenology, is to become a household word, how can we talk about it in accessible language? If the nature of fundamental concepts such as space, time, reality, probability, causality and consciousness is to be explained in terms of quantum mechanics, phenomenology, and neuroscience, how does the average citizen have a chance to understand?

I hope Todes can help. He writes: ". . . the level at which our sense of our body enters into our sense-making is such that experience shorn of that body-sense is *unintelligible*," and, as already quoted above, "All sense we can make, whether of what is actually perceivable, or possible as conceivable, draws on our sense of our [active] body, and is unintelligible without reference to it." Since we all have similar bodies, we are justified in assuming that others have similar body-senses as well, so that we can use the human body and its movements as the foundation for the construction of our worldview – confident that it will have some claim on universality.

## 4. Space according to Todes

In High School I loved Euclidean Geometry. The triumph of seeing before my eyes the steps of a successful proof gave me a little high – the conclusiveness of the proof satisfied my hunger for certainty. But for me, Euclidean geometry has one major drawback. Most proofs require the construction of "auxiliary lines", straight lines or circles that needed to be chosen cleverly and used as stepping stones, before they were erased again from the original drawing. I never found a systematic method for the choice of these lines: you just had to be clever or lucky. Going into an exam I worried whether I would be able to find them.

A year after Euclid, as we called it then, we took Analytic Geometry. Here the game is to translate all points and lines into algebraic equations, which are then solved by well established routines. No auxiliary lines here – no guessing – no need for luck. What a relief! The method requires the construction of a frame of reference, consisting of an origin and two or three mutually perpendicular axes, which serve as the fixed stage on which the game of geometry plays out. I have always thought about the three axes of that Cartesian frame of reference as being interchangeable. In three dimensions there is a subtlety about labeling the three positive directions (a property called chirality or handedness), but otherwise the identification of which one is x, y, or z is arbitrary. When I think of space, I imagine a Cartesian frame with numerical values attached to its points.

I was, therefore, surprised to learn from Todes that our bodies experience the three directions in three different ways. Our left and right sides (along the y-axis, say) are mirror images, at least for the visible parts of our bodies. Their obvious differences and similarities provide the intuitive foundations for concepts of symmetry, parity, chirality, and reversibility, as well as perennial conundrums such as "Why does a mirror reverse right and left but not up and down?" and "How do you define left and right, over the telephone to someone who doesn't know the words?" In sharp contrast, front and back (along the x axis) lack mirror symmetry. Thereby the body acquires a direction, a natural forward arrow that is missing from the y-axis. Since the description and experience of motion require the specification of a direction, Todes associates the intuition of back-to-front with motion, and motion, in turn, with time.[18] The third dimension, in the vertical direction, along the z-axis, is perceived via gravity. It, too, has a direction, but unlike the body's direction, that direction is constant and global, rather than local and changeable at will. Thus, the three axes of a Cartesian coordinate system, while mathematically identical, are differentiated by the human body.

For Todes, the perception of the spatial attribute of solid objects called extension is related to the sense of motion of the active, living body. In his critique of Hume, he writes: "Hume holds that we are aware of extension through sight and touch, but he never understands that we may become aware of extension through movement."[19] The word "may" is puzzling, because Todes then proceeds to show that body movement is actually necessary for the perception of shape, as Molyneux's question reveals. Simply "staring fixedly or resting our hand inertly on an object" do not reveal its shape or extension in three dimensions. As Todes puts it: "In movement alone we are not restricted to being passively aware of objects, in the way Hume thinks characteristic of all our experience."[20]

### 5. Time according to Todes and Mermin

An early demonstration of consilience between QBism and Phenomenology was David Mermin's solution of the problem of the NOW.[7] Before that paper, the present moment, familiar though it is to everyone, played no role in physics. Mermin showed that the principal difficulty was not the relativity of simultaneity, but the compulsive exclusion of subjectivity from science. As noted, Democritus had already suspected that this attitude would lead to frustration. With the emphasis of both QBism and phenomenology on a first person perspective, the problem of the NOW melts away. NOW, according to both Mermin and Todes, is simply the moment in time at which the anticipation and prediction of a future event turns into memory of the past experience of that event. Since each of us is capable of distinguishing between those two modes of cognition, that moment is not only well defined, but surely of surpassing importance.

Mermin re-ignited the discussion among physicists and phenomenologists about the ancient question "What is Time?" (I am, of course, not referring to sophisticated ideas like string theory, loop quantum gravity, and quantum foam, which concern the *nature* of time. This essay is about its *meaning.*) Exactly half a century earlier, Todes had come to similar conclusions by a different route.

The simple, paradigmatic story Todes tells as an illustration of my perception of time is that of myself approaching, passing, and leaving behind an object.[21] Initially the object, say a girl[22], is in front of me. My direction of motion is forward, as defined by the asymmetry of my body. After the encounter, I have *passed* the girl, and she is now in my *past*. So, Todes claims, "the front-back body distinction makes possible the *passage* of time." (He points out that the *passed-past* homonymy is no mere linguistic coincidence: The verb and the noun are both *passé* in French, and *vergangen-Vergangenheit* in German.)[23] In this way, the asymmetry of my body combines with my ability to move to produce my sense of the passage of time. Building on this humble foundation, Todes devotes additional analysis to other attributes of time, such as irreversibility, continuity, and the metaphor of flow.

### 6. Time in Popular Understanding

According to one of several different versions, Ernest Rutherford said: "A theory that you can't explain to a bartender is probably no damn good."[24] If you want to avoid offending bartenders, substitute an eleven-year-old. It turns out that the concept of the NOW according to Todes and Mermin passes this test. I know, because I was such a child.

In my book about QBism, written for the public, one chapter is devoted to Mermin's treatment of the NOW.[25] It tells the story of an experiment I performed at about age 11, when I was living in Switzerland with my family. What's remarkable is the precision with which my experience matched the scenario described by Todes to explain how we perceive time. Sitting in a train from Basel to Zurich one day, I knew that we would soon pass a fake castle which fascinated me, and later turned out to be a brewery. Anticipating the passage, I concentrated on the exact moment of closest approach, and, when it arrived, shouted: NOW. I never forgot that moment 72 years ago. I had stopped time in its course. Note the ingredients of the story. The "event" mentioned by Todes was my body passing the brewery. My motion was supplied by the train. The fixed forward orientation of my body was required by my need to concentrate on what was coming. My mental image of the brewery represented the future, until, in refreshed form, it became the past.

Decades later I repeated the experiment with groups of school and university students. The re-enactments differed from my original experiment in two details. For one thing, we experienced at first hand what Mermin called *The NOW of Many People.* In our relativistic world, such a collective event makes sense only when the participants are physically close together. Our common instant occurred when we shouted NOW in unison. As soon as we dispersed, our common NOW divided into so many personal, unrelated NOWs. Secondly, we were not moving, as I had been on the train. For bodily movement we substituted a different kind of directed, continuous change. The change required for the perception of time was provided by our collective count-down: Ten, nine, eight . . . NOW. Afterwards we discussed that NOW almost as if it were a material thing – an atom of time. What happened, according to Todes's scheme, was that we raised the bodily perception of motion from the ground floor of his metaphorical building to the upper storey of conception: we *imagined* the approach of the NOW.

Some students remembered the event years later. For me, however, the re-enactments lacked the spontaneity of the original, and soon faded from memory. What I do remember, though, is the eerie, powerful sensation of the spatial approach of the NOW. I could almost see that moment approaching, like the front of a green, Swiss locomotive. I could almost feel the vibration of the floor. It was exhilarating, and my pupils loved it.

## 7. Conclusion

Among interpretations of quantum mechanics, QBism shows with special urgency that physics, in order to offer a satisfying approach to the nature of reality, must reconnect with philosophy to reach consilience on the meaning of words such as space, time, certainty, reality, belief, truth, and probability. Phenomenology is the branch of philosophy that is currently a candidate for partnering with physics to achieve this goal.

The phenomenologist Samuel Todes suggested that the human body, with its free will, its senses, and its moveability, provides the foundation for defining and understanding all human experiences. In defense of this radical proposal, he offered a number of observations: The body is universal, unchanging, and accessible. It combines mind and matter in a way that may be difficult to disentangle, but that all of us experience from the moment of birth.

In this essay I add a modest but potentially useful observation. For the purpose of public understanding of science, considering the living human body as the origin of the frame of reference for coming to terms with the world may help to overcome the impediments imposed by the unimaginably small and large scales of most of the material universe, as well as by the complexity of modern science. Accepting even parts of Todes's phenomenology may benefit not only physicists and philosophers, but popularizers of science as well.

I would like to thank Chris Fuchs for awakening my interest in this topic, and for helpful suggestions on this essay.

[11] S. Todes, **The Human Body as Material Subject of the World** (Garland Publishing, NY, 1990)

[12] Samuel Todes**, Body and World** (MIT Press, 2001). Referred to as B&W.

[13] Maxine Sheets-Johnstone, "The Body Subject: Being True to the Truths of Experience", **The Journal of Speculative Philosophy**, Volume 34, Number 1, (Penn. State U. Press, 2020), p. 2. The abstract of this paper reads: "The aim of this essay is to flesh out the **body subject** in ways anchored in Husserlian phenomenology. In accordance with this aim, it begins with clarifications essential to veridical phenomenological accounts of experience. The specific Husserlian insights that follow these clarifications are grounded in experienced realities of animate nature, a nature Husserl consistently describes in terms of an animate organism, a **subject body**. The ensuing two sections show how and why various descriptions and claims of contemporary phenomenologists, descriptions and claims commonly anchored in Merleau-Ponty's writings on the body, fail to elucidate this **body subject**, and how and why, in bypassing Husserlian insights, the descriptions and claims are phenomenologically wayward." (My bold font.)

[14] B&W p.268

[15] Ibid. p. 88

[16] Ibid. p.100

[17] Ibid. p.263

[18] Ibid. p. 49

[19] Ibid. p. 48

[20] Ibid.

[21] Ibid.

[22] Ibid p.120

[23] Ibid. p. 300

[24] https://quotepark.com/quotes/1810067-ernest-rutherford-an-alleged-scientific-discovery-has-no-merit-unles/

[25] Hans Christian von Baeyer, **QBism—The Future of Quantum Physics** (Harvard U. Press, 2016) p.21110

10